\documentclass{aa}
\usepackage{graphicx,psfig}

\newcommand{\beqa}{\begin{eqnarray*}}
\newcommand{\eeqa}{\end{eqnarray*}}
\newcommand{\beqan}{\begin{eqnarray}}
\newcommand{\eeqan}[1]{\label{#1}\end{eqnarray}}
\newcommand{\beq}{\begin{equation}}
\newcommand{\eeq}{\end{equation}}

\makeatletter
\let\@internalcite\cite
\def\cite{\@ifstar{\citeyear}{\citefull}}
\def\citefull{\def\astroncite##1##2{##1 ##2}\@internalcite}
\def\citeyear{\def\astroncite##1##2{##2}\@internalcite}
\def\citeau{\def\astroncite##1##2{##1}\@internalcite}
\def\citen{\def\astroncite##1##2{##1 (##2)}\@internalcite}
\def\possesivcite{\def\astroncite##1##2{##1's (##2)}\@internalcite}
\def\@citex[#1]#2{\if@filesw\immediate\write\@auxout{\string\citation{#2}}\fi
  \def\@citea{}\@cite{\@for\@citeb:=#2\do
    {\@citea\def\@citea{; }\@ifundefined
       {b@\@citeb}{{\bf ?}\@warning
       {Citation `\@citeb' on page \thepage \space undefined}}%
{\csname b@\@citeb\endcsname}}}{#1}}
\def\@cite#1#2{#1\if@tempswa , #2\fi}
\def\@biblabel#1{}
\makeatother

\begin{document}

\title{The Lithium Flash}

\subtitle{Thermal instabilities generated by lithium burning in RGB stars}

\author{Ana Palacios$^{1}$ \and Corinne Charbonnel$^{1}$ \and Manuel Forestini$^{2}$}

\offprints{Corinne Charbonnel ; corinne.charbonnel@obs-mip.fr} 

\institute{
Laboratoire d'Astrophysique de Toulouse, CNRS UMR5572, OMP,
14, Av. E.Belin, 31400 Toulouse, France
\and 
Laboratoire d'Astrophysique de l'Obs. de Grenoble, 414, rue de
la Piscine, 38041 Grenoble Cedex 9, France}
\date{Received June 8, 2001 / Accepted June 23, 2001}

\abstract{
We present a scenario to explain the lithium-rich phase which occurs on the
red giant branch at the so-called bump in the luminosity function. 
The high transport coefficients required to enhance the surface lithium 
abundance are obtained in the framework of rotation-induced mixing thanks 
to the impulse of the important nuclear energy released in a lithium burning shell. 
Under certain conditions a lithium flash is triggered off. 
The enhanced mass loss rate due to the temporary increase of the stellar 
luminosity naturally accounts for a dust shell formation.
}

\maketitle

\section{The lithium-rich RGB stars}
Since its discovery by Wallerstein \& Sneden (1982), 
various scenarii have been proposed to explain the lithium-rich giant phenomenon. 
Some call for external causes, like the engulfing of nova ejecta or of a planet 
(Brown et al. 1989; Gratton \& D'Antona 1989; Siess \& Livio 1999). 
Others refer to fresh lithium production by the Cameron \& Fowler (1971, hereafter CF71) 
mechanism associated to a transport process in the deep radiative layers 
of the red giants (Fekel \& Balachandran 1993; de la Reza et al. 1997; 
Sackmann \& Boothroyd 1999; Jasniewicz et al. 1999; 
Charbonnel \& Balachandran 2000, hereafter CB00; 
Denissenkov \& Weiss 2000, hereafter DW00). 
The latter explanation is sustained by the depletion of Be in the atmosphere 
of the lithium-rich giants (Castilho et al. 1999), 
and by the location of these objects on the Hertzsprung-Russel diagram (CB00). 
Indeed Hipparcos parallaxes allowed the localisation of the short episode 
of lithium production at the bump of the luminosity function on the red giant
branch (RGB). 
CB00 thus concluded that the lithium-rich RGB phase was a precursor 
to the mixing process which leads to other abundance anomalies, like 
the low carbon isotopic ratios seen in $\sim 96\%$ of the 
low-mass stars ($\leq$2.2 - 2.5M$_{\odot}$; Charbonnel \& Do Nascimento 1998). 
This brought very tight constraints on the underlying mixing process 
and in particular on its extreme rapidity. 
It put in difficulty models where high lithium abundances 
can be reached at any time between the RGB bump and tip
(Sackmann \& Boothroyd 1999, DW00).

In this paper we propose a scenario which accounts for the high 
transport coefficients required to enhance the surface lithium abundance 
of RGB stars in the framework of rotation-induced mixing (\S 2). 
We show that under certain conditions a thermal instability 
can occur, triggered by the nuclear burning of the freshly synthetized 
lithium in a lithium burning shell (LiBS) external to the classical hydrogen 
burning shell (HBS; \S 3)
\footnote{Our scenario is different from the previous works which explored the 
possible occurrence of shell flashes on the RGB 
and which considered only the reactions of the CNO cycle 
(Bolton \& Eggleton 1973, Dearborn et al. 1975, Von Rudloff et al. 1988). 
It also differs from the model developed by Fujimoto et al. (1999) 
to explain Mg and Al anomalies in globular cluster red giants.
These authors indeed proposed that a flash is triggered off deeper in the HBS 
due to the inward mixing of H down into the degenerate He core. }. 
The enhanced mass loss related to the increase of the stellar luminosity 
during the lithium-rich phase can explain the circumstellar dust shells 
observed around some of these objects. 

\section{The scenario}
\subsection{Rotation induced mixing}

Observational evidence has accumulated on the existence of an extra-mixing 
process which becomes efficient on the RGB when the low-mass stars 
reach the luminosity function bump
(Brown 1987; Gilroy \& Brown 1991; Pilachowski et al. 1993; 
Charbonnel 1994; Charbonnel et al. 1998; Gratton et al. 2000).
At this point the outwardly-moving HBS crosses 
the mean molecular weight discontinuity created by the first dredge-up. 
As shown in Charbonnel (1995) and Charbonnel et al. (1998), an extra-mixing 
process related to rotation and previously inhibited by a 
strong molecular weight gradient can then easily connect the 
convective envelope to the deeper layers where the nucleosynthesis 
occurs and consequently lower the carbon isotopic ratio.
Rotation was also invoked to account for C,N,O and Na anomalies and $^3$He and $^7$Li 
destruction in low-mass and low-metallicity RGB stars 
(Sweigart \& Mengel 1979; Charbonnel 1995; Denissenkov \& Tout 2000, hereafter DT00). 

Recently, DW00 investigated the effect of rotation-induced mixing on the surface 
lithium abundance using prescriptions based on Zahn and Maeder developments 
(Zahn 1992, Maeder \& Zahn 1998). They used the corresponding transport coefficients 
D$_{\rm R}$ derived by DT00 for globular cluster low-mass giants.   
They found that these values of D$_{\rm R}$ were too low 
to lead to lithium enrichment and invoked the engulfing of a planet in order to 
trigger and dope the transport process. While this combined scenario is 
rather attractive, it is harmed by the fact that it can happen at any time
on the RGB while lithium-rich giants with solar metallicity discovered up to now 
do stand by the bump (CB00).
However what was not investigated by DT00 and DW00 is the structural 
response of the star to the extra-mixing. Their computations of the 
transport coefficients and of the abundance variations are indeed done 
in a post-processing approach where standard stellar structures are 
used as background models. 
We propose here that the {\em structural response to the mixing}  
is actually the {\em cause of the increase of D$_{\rm R}$} which is necessary to
enhance the surface lithium abundance.

\subsection{Impact of the nuclear energy released by lithium burning}

\begin{figure}
\resizebox{\hsize}{!}{\includegraphics{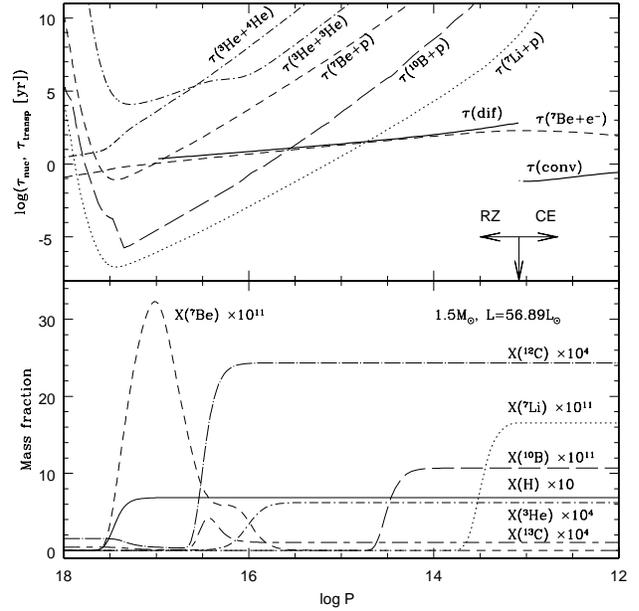}}
\caption{1.5M$_{\odot}$, Z$_{\odot}$ star at the RGB bump - 
{\it (top)} Lifetime of $^7$Li, $^7$Be, $^{10}$B and $^3$He, and convective 
and diffusive time scales 
(bold lines; $\tau_{\rm (dif)}$ is shown for a transport coefficient constant 
and equal to 10$^{11}$cm$^2$sec$^{-1}$). The transition between the 
convective enveloppe and the radiative zone is indicated by the arrows.
{\it (bottom)} Abundance profiles of the relevant elements (in mass fraction) in 
the same region}
\label{fig1PalaciosetalDf113}
\end{figure}

Fig.~\ref{fig1PalaciosetalDf113} shows the abundance profiles of the 
relevant elements around the HBS in a 1.5M$_{\odot}$, Z$_{\odot}$ star 
(typical of the lithium-rich RGB giants; see CB00) which is just passing 
through the RGB bump and for which the extra-mixing is still inactive. 
Since the end of the first dredge-up the mass of the core has grown 
and the convective envelope has retreated. 
At the bump the external wing of the $^7$Be peak crosses the 
molecular weight discontinuity before the $^{13}$C peak does.  
When it is connected with the convective envelope by the extra-mixing process, 
$^7$Be produced via $^3$He($\alpha,\gamma)^7$Be starts to diffuse outwards.  
Depending on the mixing timescale $\tau$(dif), the CF71 mechanism may or may not operate 
in this radiative region. According to DW00 transport coefficients D$_{\rm R}$ 
$\sim 10^{11}$cm$^2$sec$^{-1}$ are needed to increase the surface lithium abundance. 
This can be easily understood by looking at Fig.~\ref{fig1PalaciosetalDf113}
where the lifetimes of the various elements are compared with 
$\tau$(dif) computed for this value of D$_{\rm R}$. 
In particular, we then have $\tau$(dif) $\simeq \tau(^7$Be+e$^-$) everywhere below
the convective envelope. Higher D$_{\rm R}$ 
are thus necessary for the freshly synthetized $^7$Li to survive, 
$^7$Be being transported rapidly to cooler regions before it decays. 
In other words, if we assume that the extra-mixing is related to rotation and 
starts with the low values obtained by DT00 ($10^8$ to 10$^{10}$cm$^2$sec$^{-1}$), 
then at the beginning of this phase $\tau$(dif) $> \tau(^7$Be+e$^-$) : 
$^7$Be diffuses outwards but decays in regions where $^7$Li is rapidly destroyed 
by proton capture. 

\begin{figure}
\resizebox{\hsize}{!}{\includegraphics{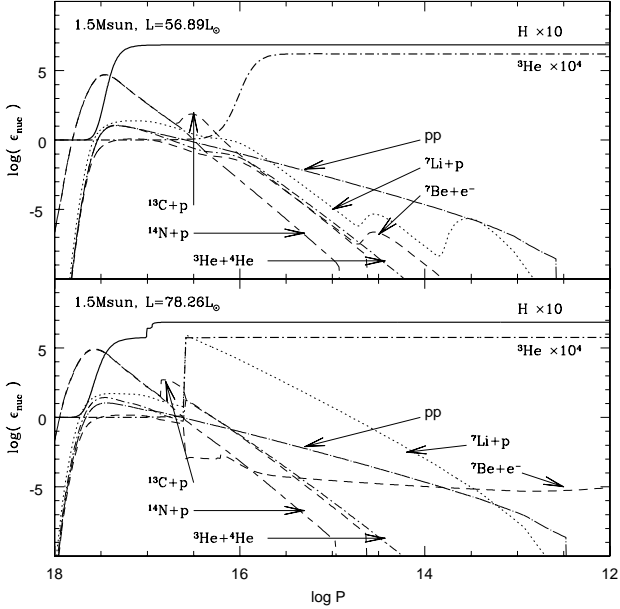}}
\caption{Energy generation rate [erg.g$^{-1}$.sec$^{-1}$] for various 
nuclear reactions in the 1.5M$_{\odot}$, Z$_{\odot}$ star at the RGB bump.
The H and $^3$He abundance profiles (in mass fraction) are also shown - 
{\it (top)} Model at the bump, just before the start of the extra-mixing.
{\it (bottom)} Model with mixing. $\epsilon_{\rm nuc}(^7$Li+p) is 
now the main contributor to the energy released in the external wing 
of the hydrogen burning shell. 
$\epsilon_{\rm nuc}(3$He+$^4$He) is also significantly increased 
as well as $\epsilon_{\rm nuc}(^7$Be+e$^-$) in the more external layers}
\label{fig2PalaciosetalDf113}
\end{figure}

In order to test the reaction of the stellar structure in such a situation, 
we computed an evolutionary sequence for a 1.5M$_{\odot}$, Z$_{\odot}$ model assuming 
that the mixing starts at the bump and using a parametrized D$_{\rm R}$
that we take first of the order of 10$^9$cm$^2$sec$^{-1}$ 
between the base of the convective envelope and the external wing of the $^7$Be peak. 
Details of the stellar code and of the input physics will be given in Palacios et
al. (2001).
The energy generation rates for the main nuclear reactions are shown in 
Fig.~\ref{fig2PalaciosetalDf113}. 
Before the mixing starts $^{14}$N(p,$\gamma)^{15}$O dominates inside the HBS.
Then as evolution proceeds with the initially low D$_{\rm R}$, 
$^7$Be diffuses outwards and decays in the radiative layer where $^7$Li is rapidly 
destroyed by proton capture.
In this thin region the energy released by $^7$Li(p,$\alpha)\alpha$ 
increases significantly.
This reaction becomes actually the dominant one.
In addition to the HBS the star has now a {\em very thin lithium burning shell (LiBS)}.
In the region where the mixing acts the contributions of $^7$Be(e$^-,\nu)^7$Li and
$^3$He($^4$He,$\gamma)^7$Be also increase though to a lesser extent. 
This change in the energy production 
leads to an increase of the stellar luminosity and radius which are rapidly 
shifted from $\sim$57 to $\sim$78L$_{\odot}$ and from $\sim$13 to 
$\sim$16.5R$_{\odot}$ respectively in the case presented here.
Consequently the mass loss rate (given by Reimers' 1975 approximation in our 
computations) is almost doubled. One thus expects to see the
formation of a dust shell around the star as suggested by the IRAS fluxes 
around some of the lithium-rich objects (de la Reza et al. 1996, 1997).

In Zahn and Maeder developments for the transport of matter and angular momentum, 
the effective diffusivity is proportional to [r U(r)]$^2$, 
the vertical component of the meridional velocity being itself 
proportional to [E$_{\Omega}+$E$_{\mu}$]. 
E$_{\Omega}$ and E$_{\mu}$ depend respectively on the velocity profile and 
on the chemical inhomogeneities along isobars (see e.g. Maeder \& Zahn 1998 for 
their complete expressions).
But what is important for the present discussion is their dependence on the local 
rates of nuclear and gravothermal energies. 
Here we thus propose that the {\em apparition of the LiBS and the significant 
release of $\epsilon_{\rm nuc}(^7$Li+p) do highly enhance the meridional 
velocity and the associated transport coefficient} as required 
for the surface lithium abundance to increase.

\section{Thermal instabilities generated by lithium burning}
In order to mimic this situation we raised the value of D$_{\rm R}$
(up to $10^{12.5}$) as $\epsilon_{\rm nuc}(^7$Li+p) increased. 
The mixing depth was kept constant. 
We are aware of the crudeness of our approach, and of the necessity to 
follow in a self-consistent way the evolution of the transport of matter and 
of angular momentum since the early phases of evolution to get the mixing 
characteristics (see e.g. Charbonnel \& Talon 1999). 
However the model used here already gives the main trends of the structural 
response of the star to such a mixing. 

\begin{figure}
\resizebox{\hsize}{!}{\includegraphics{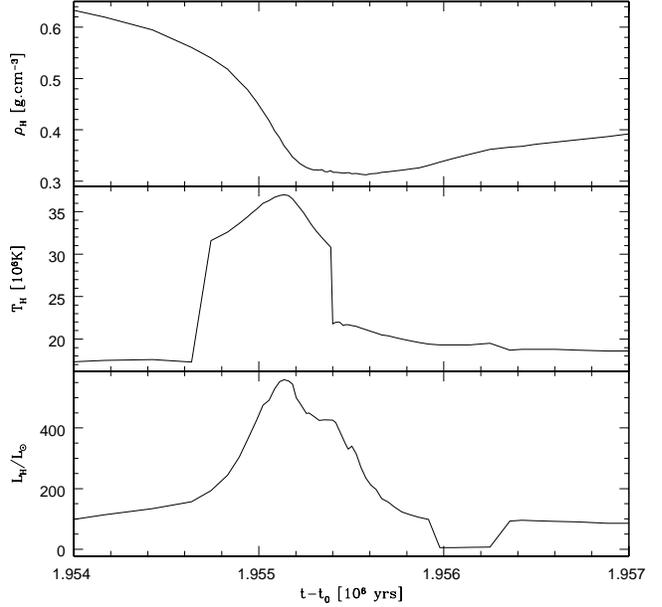}}
\caption{Characteristics of the lithium flash in the 1.5M$_{\odot}$, 
Z$_{\odot}$ model in terms of time variations since the beginning of the bump 
(t$_0$) of the density {\it (upper panel)}, temperature {\it (middle panel)} 
and nuclear luminosity {\it (lower panel)} at the base of the LiBS}
\label{fig3PalaciosetalDf113}
\end{figure}

In particular, a {\em thermal instability fed by the lithium burning}, or 
{\em lithium flash}, 
could develop with the characteristics shown in Fig.3.  
In the thin region where the mixing acts, $\epsilon_{\rm nuc}(^7$Li+p) strongly 
increases  (up to $\sim 3.2\times 10^6$ erg g$^{-1}$ sec$^{-1}$ in the model presented 
here while in a standard HBS $\epsilon_{\rm nuc}$(CN) is of the order of 
$2 \times 10^5$ erg g$^{-1}$ sec$^{-1}$ at this evolutionary stage). 
This leads to a fast increase of the local luminosity.
While the LiBS tries to get rid of the nuclear energy excess its density drops.  
However its expansion is not sufficient to lift the overlaying regions and the 
corresponding pressure decrease is rather small. 
The temperature thus keeps increasing, as well as $\epsilon_{\rm nuc}(^7$Li+p). 
Consequently in the thin LiBS the heating timescale 
(c$_{\rm P}$T/$\epsilon_{\rm nuc}(^7$Li+p)) becomes shorter than the 
timescale of heat diffusion 
(c$_{\rm P}$T H$_{\rm p}^2 \kappa \rho^2 / (1-\beta)$cP). 
The thermal instability is then strong enough to require a convective transport 
of the energy. 
As a consequence $^7$Be can be instantaneously transported in the external 
convective envelope. 
This results in a sudden increase of the surface lithium abundance. 
In the case presented here the lithium mass fraction in the convective envelope 
was increased by a factor 3 compared to its post dredge-up value. 
This result is of course dependent on the apparition and on the strength of 
the lithium flash, which itself depends on the way the transport coefficient evolves 
as $\epsilon_{\rm nuc}(^7$Li+p) is dumped in the thin layer between the Be peak and
the convective envelope. 
Quantitative conclusions require a self-consistent treatment of the whole process.  

A response of D$_{\rm R}$ to $\epsilon_{\rm nuc}(^7$Li+p) different from the one
simulated here could lead to a situation where both the stellar luminosity sustained by 
$\epsilon_{\rm nuc}(^7$Li+p) and the surface lithium abundance increase 
more drastically before the lithium flash. 
In this case lithium-rich stars with very high luminosity could be produced 
during a very short phase. 
This could explain the rare occurrence of very bright lithium-rich giants such 
as the two stars discovered in the globular clusters M3 (Kraft et al. 1999) 
and NGC 362 (Smith et al. 1999) which stand at much higher luminosity. 

At this stage one can however speculate on the way the lithium-rich phase ends. 
Indeed when the convective instability develops, it erases the mean molecular 
weight gradient which was still protecting the deeper layers from the mixing. 
The transport process is then free to proceed closer to the core. 
Once it reaches the regions where $^{12}$C is nuclearly converted into $^{13}$C 
(as is observed to occur in low-mass stars after the bump) the surface material
is exposed to temperatures higher than lithium can withstand. 
The freshly synthesized lithium is then destroyed as the surface carbon isotopic 
ratio decreases and the lithium-rich phase ends. 
This is consistent with the available data on $^{12}$C/$^{13}$C in the 
lithium-rich RGB stars : those with the highest lithium abundance still have a 
standard (i.e. post-dredge-up) carbon isotopic value, but no star 
retains its peak lithium abundance once its carbon isotopic value dips (CB00).

In order to estimate the timescale of our process let us define 
the beginning and the end of the lithium-rich phase as respectively the moments where
the surface lithium abundance increases above its post-dredge up value and where the 
carbon isotopic ratio decreases below its post-dredge up value. 
While the duration of the lithium flash itself is quite short in the model presented 
here ($\sim 2 \times 10^4$ yrs; see Fig.3), the lithium-rich phase so-defined lasts 
for $\sim 60 \%$ (i.e. $\sim 6 \times 10^6$ yrs) of the classical bump duration. 
This has to be compared with the number of stars located in the bump region and
which still present a post-dredge up carbon isotopic ratio. 
Among the sample of stars with Hipparcos parallaxes and carbon isotopic ratio 
determinations used in Charbonnel \& do Nascimento (1998), $\sim 35 \%$ of the 
49 objects located close to the bump do satisfy this criterion. 
We consider that this comparison is very encouraging in view of the simplifications 
we made in the present model and of the limited observational sample. 
This preliminary estimation indicates that many lithium-rich stars 
should be discovered by observational programs focussing on the bump region. 

\section{Conclusions}

We propose a scenario to explain the lithium-rich phase at the 
RGB bump in the framework of rotation-induced mixing.  
When the external wing of the $^7$Be is connected to the convective envelope the
extra-mixing timescale is at first such that the transported $^7$Be decays 
in the region where $^7$Li is destroyed by p-capture. 
A lithium burning shell appears.
The meridional circulation and the corresponding transport coefficient 
then increase due to the $\epsilon_{\rm nuc}(^7$Li+p) burst. 
$^7$Li(p,$\alpha)\alpha$ becomes the dominant reaction leading to an increase of the 
local temperature and of both the local and total luminosities. 
A lithium flash terminates this phase. 
As the convective instability develops it erases the molecular weight gradient 
and the mixing is free to proceed deeper. Both the lithium abundance and the 
carbon isotopic ratio then drop at the stellar surface.
During the whole sequence the temporary increase of the stellar luminosity 
causes an enhanced mass loss rate which naturally accounts for the dust shell suggested 
by the far-infrared color excesses measured for some lithium-rich objects from 
IRAS fluxes.
However the contribution of these stars to the lithium enrichment of the Galaxy
should be very modest.

The treatment of the transport coefficients used in the present simulation 
is rudimentary and serves only to outline our scenario. 
In particular the apparition and the strength of 
the lithium flash depend on the way the transport coefficient evolves  
as $\epsilon_{\rm nuc}(^7$Li+p) is dumped in the thin layer 
between the Be peak and the convective envelope. 
This has to be followed in a model where the structure and the transport of 
angular momentum and of the chemical species are treated self-consistently. 
This will be presented in a forthcoming paper where consequences on the luminosity
function will also be investigated (Palacios et al. 2001). 

\begin{acknowledgements}
We dedicate this paper to the lithium babies No\'e and Nishant.
We acknowledge support from the Programme National de Physique Stellaire and the
Programme National Galaxies.
\end{acknowledgements}

\appendix


\begin{thebibliography}{}
\bibitem[1973]{}Bolton A.J.C., Eggleton P.P., 1973, A\&A 24, 429
\bibitem[1987]{}Brown J.A., 1987, ApJ 317, 701
\bibitem[1989]{}Brown J.A., Sneden C., Lambert D.L., Dutchover E.Jr, 
1989, ApJsupplt 71, 293
\bibitem[1971]{}Cameron A.G., Fowler W.A., 1971, ApJ 164, 111, CF71
\bibitem[1999]{}Castilho B.V., Spite F., Barbuy, B., Spite M., de Medeiros 
J.R., Gregorio-Hetem J., 1999, A\&A 345, 249
\bibitem[1995]{}Charbonnel C., 1994, A\&A 282, 811
\bibitem[1995]{}Charbonnel C., 1995, ApJ 453, L41
\bibitem[2000]{}Charbonnel C., Balachandran S.C., 2000, A\&A 359, 563, CB00
\bibitem[1995]{}Charbonnel C., Brown J.A., Wallerstein G., 1998, A\&A 332, 204
\bibitem[1998]{}Charbonnel C., do Nascimento J.D.J., 1998, A\&A 336, 915
\bibitem[2000]{}Charbonnel C., Talon S., 1999, A\&A 351, 635
\bibitem[1975]{}Dearborn D.D., Bolton A.J.C., Eggleton P.P., 1975, MNRAS 170, 7
\bibitem[1997]{}de la Reza R., Drake N.A., da Silva L., 1996, ApJ 456, L115
\bibitem[1997]{}de la Reza R., Drake N.A., da Silva L., Torres C.A.O., 
Martin E.L., 1997, ApJ 482, L77
\bibitem[2000]{}Denissenkov P.A., Tout C.A., 2000, MNRAS 316, 395, DT00
\bibitem[2000]{}Denissenkov P.A., Weiss A., 2000, A\&A 358, L49, DW00 
\bibitem[1993]{}Fekel F.C., Balachandran S., ApJ 403, 708
\bibitem[1999]{}Fujimoto M.Y., Aikawa M., Kato K., 1999, ApJ 519, 733
\bibitem[1991]{}Gilroy K.K., Brown J.A., 1991, ApJ 371, 578
\bibitem[1989]{}Gratton R.G., D'Antona F., 1989, A\&A 215, 66
\bibitem[2000]{}Gratton R.G., Sneden C., Carretta E., Bragaglia A., 2000, 
A\&A 354, 169
\bibitem[1999]{}Jasniewicz G., Parthasarathy M., de Laverny P., Thevenin F., 
1999, A\&A 342, 831
\bibitem[1999]{}Kraft R.P., Peterson R.C., Guhathakurta P., et al., 1999, 
ApJ 518, L53
\bibitem[1999]{}Maeder A., Zahn J.P., 1998, A\&A 334, 1000
\bibitem[2001]{}Palacios A., Charbonnel C., Forestini M., Talon S., 2001, A\&A, in preparation
\bibitem[1993]{}Pilachowski C.A., Sneden C., Booth J., 1993, ApJ 407, 699
\bibitem[1975]{}Reimers D., 1975, Mem. Soc. Roy. Sci. Li\`ege 8, 369
\bibitem[1999]{}Sackmann I.J., Boothroyd A.I., 1999, ApJ 510, 217
\bibitem[1999]{}Siess L., Livio M., 1999, MNRAS 308, 1133
\bibitem[1999]{}Smith V.V., Shetrone M.D., Keane M.J., 1999, ApJ 516, L73
\bibitem[1979]{}Sweigart A.V., Mengel K.G., 1979, ApJ 229, 624
\bibitem[1988]{}Von Rudloff I.R., VandenBerg D.A., Hartwick F.D.A., 1988, ApJ 324, 840
\bibitem[1982]{}Wallerstein G., Sneden C., 1982, ApJ 255, 577
\bibitem[1992]{}Zahn J.P., 1992, A\&A 265, 115
\end{thebibliography}
\end{document}